\documentclass[aps,10pt,reprint,groupedaddress,superscriptaddress,nofootinbib]{revtex4-2}

\usepackage[usenames,dvipsnames]{xcolor}
\usepackage{amssymb}
\usepackage{graphicx}
\usepackage{amsmath}
\usepackage{dsfont}
\usepackage{txfonts}
\usepackage[bookmarks=true,colorlinks,citecolor=blue,urlcolor=blue]{hyperref}
\usepackage{braket}
\usepackage{babel}
\usepackage{blindtext}
\usepackage{physics}
\usepackage{enumitem}
\usepackage[normalem]{ulem}

\newcommand{\be}{\begin{equation}}
\newcommand{\ee}{\end{equation}}
\newcommand{\fig}[1]{Fig.\thinspace{}\ref{#1}}
\newcommand{\fc}[1]{({#1})}
\newcommand{\figc}[2]{Fig.\thinspace{}\ref{#1}\thinspace{}\fc{#2}}

\begin{document}

\newcommand{\TUM}{\affiliation{Technical University of Munich, TUM School of Natural Sciences, Physics Department, 85748 Garching, Germany}}
\newcommand{\MCQST}{\affiliation{Munich Center for Quantum Science and Technology (MCQST), Schellingstr. 4, 80799 M{\"u}nchen, Germany}}
\newcommand{\LMU}{\affiliation{Fakult\"at f\"ur Physik, Ludwig-Maximilians-Universit\"at, 80799 Munich, Germany}}
\newcommand{\MPQ}{\affiliation{Max-Planck-Institut f\"ur Quantenoptik, 85748 Garching, Germany}}

\def\papertitle{{Preparing and Analyzing Solitons in the sine-Gordon Model with Quantum Gas Microscopes}}
\title{\papertitle}

\author{Elisabeth Wybo} \TUM \MCQST
\author{Alvise Bastianello} \TUM \MCQST
\author{Monika Aidelsburger} \MCQST \LMU 
\author{Immanuel Bloch} \MCQST \LMU \MPQ 
\author{Michael Knap} \TUM \MCQST

\date{\today}

\begin{abstract}
The sine-Gordon model emerges as a low-energy theory in a plethora of quantum many-body systems. Here, we theoretically investigate tunnel-coupled Bose-Hubbard chains with strong repulsive interactions as a realization of the sine-Gordon model deep in the quantum regime. 
We propose protocols for quantum gas microscopes of ultracold atoms to prepare and analyze solitons, that are the fundamental topological excitations of the emergent sine-Gordon theory. With numerical simulations based on matrix product states we characterize the preparation and detection protocols and discuss the experimental requirements. 
\end{abstract}

\maketitle

\section{Introduction}

Universality forms one of the pillars for the classification of quantum phases of matter. Upon coarse-graining, microscopic details become irrelevant and only symmetry and topology determine the essential properties. Seemingly different looking microscopic systems can then be described by the same set of collective degrees of freedom that are captured by the same emergent effective field theory.
A prominent example is the relativistic sine-Gordon field theory \cite{coleman1975,FADDEEV19781} which emerges as the low-energy description of various physical systems, including among others the massive Thirring model~\cite{coleman1975,Cirac2010}, the Coulomb gas~\cite{Stu1978}, spin chains~\cite{Affleck1999, Zvyagin2004, Umegaki2009}, bosonic and fermionic Hubbard models~\cite{Essler2005, giamarchi2003quantum}, and  circuit quantum electrodynamics~\cite{Roy2019, Roy2021}; and thus is of central interest for a multifaceted community.

In the strongly interacting regime, the sine-Gordon model possesses a complex quasiparticle spectrum, consisting of solitons, that are massive topological excitations, and breathers, that are bound states of these solitons. Furthermore, the sine-Gordon model is a renowned example of an integrable field theory \cite{Smirnov1992,Zamolodchikov1979}, which implies infinitely long-lived quasiparticles, unconventional relaxation dynamics~\cite{Calabrese_2016}, and unconventional transport \cite{Bertini2021, specialissueGHD}. Elegant analytical methods can be used to obtain exact results which have improved our understanding of the model \cite{Smirnov1992,Zamolodchikov1979,Bertini2014,Kormos2016,Cubero2017,Kukuljan2018,Rylands2019,Bertini2019,Kukuljan2020}.
Despite these advances from integrability, analytical predictions for correlation functions and the full-counting statistics of many observables, both in equilibrium and out of equilibrium, are very difficult to access.
Moreover, it is pertinent to understand in which dynamical regimes the effective sine-Gordon field theory is realized in one of the microsocopic models described above~\cite{Wybo2022}. This can lead to the development of highly-tunable sine-Gordon quantum simulators, that are for instance based on ultracold atoms in optical lattices, Rydberg atoms, or trapped ions.

Pioneering theoretical work~\cite{Gritsev2007,Gritsev2007a} motivated the experimental realization of the sine-Gordon model with tunnel-coupled, one-dimensional quasi-condensates~\cite{Schweigler2017,Zache2020,Pigneur2018}. In these atom chip implementations, the sine-Gordon mass scale can be tuned and correlation functions can be efficiently characterized by matter-wave interferometry \cite{Schumm2005,Hofferberth2007,Langen2015,Schweigler2017,Rauer2018,Yuri2018}, but the interactions in the one-dimensional gases have been restricted to be rather weak. As a consequence, the emergent sine-Gordon field theory is approximately semiclasscial~\cite{Schweigler2017,Blakie2008,DeLuca2016}. From this perspective, tunnel-coupled optical lattice systems can be very convenient: they allow the kinetic energy to be quenched and, hence, the effective interactions to be enhanced. 
Although in a previous theoretical work \cite{Wybo2022} we have shown that the sine-Gordon model describes the low-energy physics of two coupled spin chains very well, identifying an experimental realization of such a setup and developing the measurement protocols to extract the information of the emergent field theory remain unaddressed challenges that motivate our present work.

\begin{figure*}
    \centering
    \includegraphics[width=0.9 \textwidth]{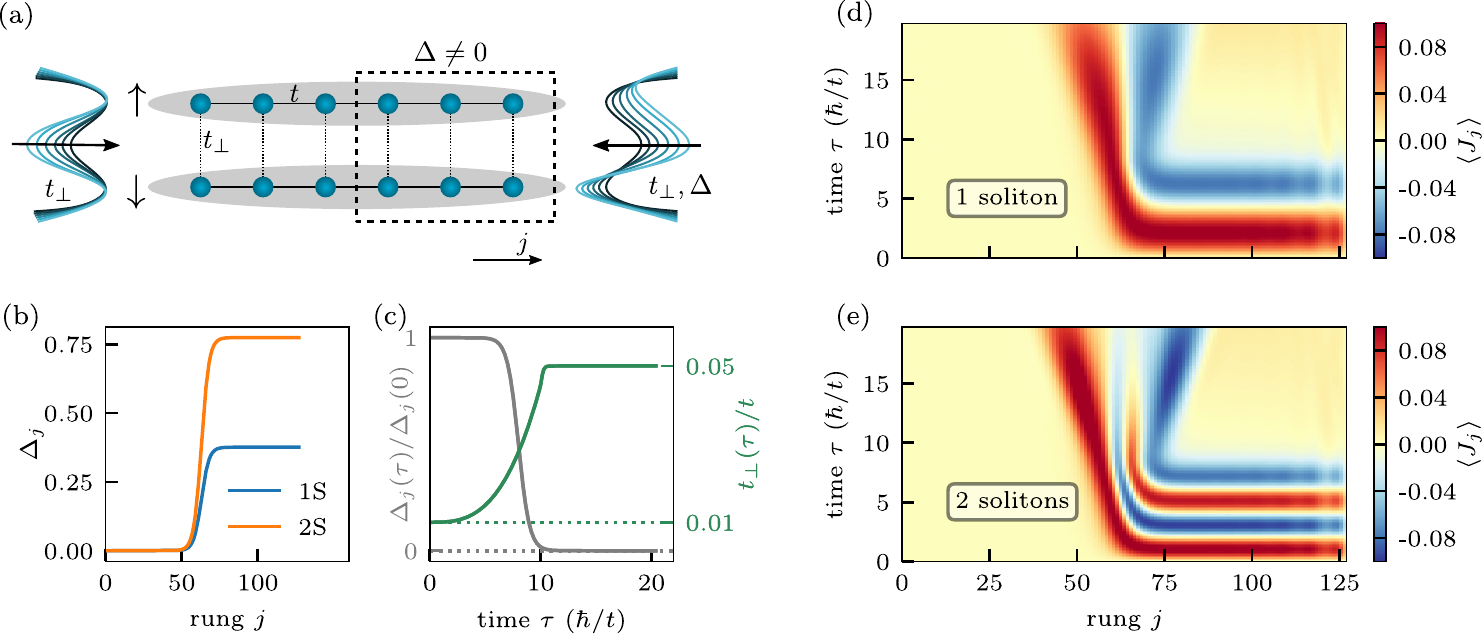}
    \caption{\textbf{Creation of solitons in coupled Bose-Hubbard chains.} (a) A soliton can be created by applying a chemical-potential gradient between the two chains in one half of the system $ \sum_j \Delta_j(\tau) (n_{j\uparrow}-n_{j\downarrow})$, as indicated by the dashed region. Ramping up the tunnel coupling $t_{\perp}$ simultaneously, which decreases the barrier on the rung, enables us to further stabilize the phase imprinting process. (b) The spatial profile of the applied chemical potential gradient $\Delta_j$ for the preparation of one (1S) and two (2S) solitons. (c) The time dependence of the chemical potential difference $\Delta_j$ and tunnel coupling $t_{\perp}$. (d,e) The resulting profiles of the mean local current $\langle J_j\rangle$ suggest the creation of one (d) and two (e) solitons, respectively. {The time-evolution is performed with the TEBD algorithm for bond dimension $\chi=1000$ and $L=128$ rungs. }  }
    \label{fig:sol_curr}
\end{figure*}

Here, we show how solitons, which are the fundamental excitations of the sine-Gordon model, can be created in a setting of ultracold bosons trapped in an optical lattice [see Fig.~\ref{fig:sol_curr}(a)]. The solitons are most directly observed when measuring the current between the chains by first applying beam splitter operations and then taking snapshots with a quantum gas microscope. Using similar beam-splitter operations, we furthermore show how the full counting statistics of the topological charge, that counts the number of excited
solitons, can be obtained. We use numerical simulations based on matrix product states, to assess the experimental requirements of the proposed protocols.

The paper is organized as follows: in Section \ref{sec_setup} we briefly introduce the sine-Gordon model and its realization using coupled one-dimensional Bose-Hubbard chains. 
The creation of a soliton is discussed in Section \ref{sec_solitons}. In Section \ref{sec_soliton_detection} we present a detection method to capture the topological charge of solitons. The requirements for the experimental preparation of a low-energy state of the sine-Gordon field theory is discussed in Section \ref{sec:exp_re}. Conclusions are presented in Section \ref{sec_conclusions} and the appendixes contain a couple of technical details.

\section{Tunnel-coupled Bose-Hubbard chains}
\label{sec_setup}

The sine-Gordon model emerges as the low-energy description of two tunnel-coupled, interacting, one-dimensional systems.  
Here, we consider two identical Bose-Hubbard chains with Hamiltonians $H^{\text{BH}}_{\uparrow}$ and $H^{\text{BH}}_{\downarrow}$, coupled with a tunneling term $H_{\perp}$.
The dynamics is therefore set by the total Hamiltonian $
H=H^{\text{BH}}_{\uparrow} + H^{\text{BH}}_{\downarrow} +   H_{\perp}$. Here,
\be
    H_{\alpha}^{\text{BH}} = - t \sum_{j=0}^{L-2}  (b^{\dag}_{j\alpha} b_{j+1\alpha} + \mathrm{h.c.}) 
    + \frac{U}{2} \sum_{j=0}^{L-1} n_{j\alpha}(n_{j\alpha}-1) \, ,
\ee
where $b_{j\alpha}$ ($b_{j\alpha}^\dagger$) annihilates (creates) a boson on rung $j$ in chain $\alpha=\{\uparrow,\downarrow\}$ and $n_{j\alpha}=b^\dagger_{j\alpha}b_{j\alpha}$ is the density. The tunnel coupling Hamiltonian reads
\be
    H_{\perp} = -t_{\perp} \sum_{j=0}^{L-1} (b^{\dag}_{j\uparrow} b_{j\downarrow} + \mathrm{h.c.}). 
\ee
We use open boundary conditions throughout our work.
The derivation of the low-energy description of the coupled wires closely follows the original proposal \cite{Gritsev2007}.
We consider each of the chains to be at non-integer filling to realize a superfluid state for arbitrary values of the hopping $t$ and the interaction strength $U$. 

In the absence of inter-chain coupling $t_\perp=0$, the low-energy behavior is obtained by bosonization. 
One first introduces the phase field $\phi_\alpha(x)$, and its conjugate field $\Pi_\alpha(x)$. These two fields are related to the microscopic operators as $b^{\dagger}_{j,\alpha} \simeq  \gamma e^{i\phi_\alpha(x)}$ and  as $n_{j,\alpha}\simeq n + \Pi_\alpha(x)$, respectively, where $\gamma$ is a non-universal prefactor and $n$ is the average density. Within bosonization, the two one-dimensional chains are governed by the Luttinger-Liquid Hamiltonian
 \be
H^{\mathrm{LL}}_{\alpha}=\int \dd x \frac{\hbar v_s}{2} \left(\frac{\pi}{K}  \Pi_{\alpha}^2(x)+\frac{K}{\pi}(\partial_x \phi_{\alpha}(x))^2\right)\, ,
\ee
where, $K$ is the Luttinger parameter and $v_s$ is the sound velocity. The Luttinger parameter is larger than one, $K\geq 1$, for repulsive interactions $U>0$ \cite{giamarchi2003quantum}, and approaches $K=1$ in the limit of infinite repulsive interactions $U/t\to \infty$. For weak tunneling $t_\perp\ll t$, the interchain coupling can be reintroduced perturbatively~\cite{Gritsev2007}. Upon bosonizing the transverse Hamiltonian reads as $H_\perp=-2|\gamma|^2 t_\perp\int \dd x  \cos(\phi_\uparrow-\phi_\downarrow)$.
Performing a rotation to symmetric $\psi=\phi_{\uparrow}+\phi_{\downarrow}$ and antisymmetric $\phi=\phi_{\uparrow}-\phi_{\downarrow}$ degrees of freedom, results in the explicit decoupling of the two sectors. The symmetric combination is governed by a gapless Luttinger Liquid, and the sine-Gordon field theory \eqref{eq_H_SG} emerges for the relative degree of freedom $\phi$:

\be\label{eq_H_SG}
    H^{\text{SG}} = \int \dd x \, \frac{\hbar v_s}{2} \left(\frac{2\pi}{K} \Pi^2+\frac{K}{2\pi}(\partial_x \phi)^2\right) 
-2 t_{\perp}  |\gamma|^2\cos\phi\, .
\ee

The integrability of the sine-Gordon model allows for an exact determination of its particle content, together with an analytical expression for the two-body scattering matrix \cite{Zamolodchikov1979}.
The fundamental excitations are topological solitons and anti-solitons, for which the phase field $\phi$ interpolates  between the degenerate minima of the $\cos\phi-$interaction. Therefore, the phase field $\phi$ winds up by $2\pi$ when traversing a soliton. These quasiparticles have a relativistic dispersion law $E_s(k)=\sqrt{v_s^4M^2+v_s^2k^2}$. The soliton mass $M$ has a complicated dependence on the bare parameters of the Hamiltonian~\cite{Zamolodchikov1995}, but scales with the transverse hopping $t_\perp$ in a rather simple manner $M\propto t_\perp^{\frac{2K}{4K-1}}$. 

Depending on the interaction $K$, solitons and anti-solitons can form bound states of infinite lifetime, which are called breathers. The interaction $K$ sets the number of breathers in the excitation spectrum, $N=\lfloor 4K-1 \rfloor$, and their mass as $M_{B_n}=2M \sin\left(\frac{\pi}{2}\frac{n}{4K-1}\right)$. Crucially, when $K$ is increased the relative mass difference between two consecutive breathers is reduced: in the limit of large $K$, or weak interactions, the breather's mass spectrum collapses to a continuum and the quantum model is well-approximated by the classical sine-Gordon theory \cite{Blakie2008,faddeev2007hamiltonian,Chung1989}. In contrast, the deep quantum regime of the field theory is realized when only few breathers are present in the spectrum. In tunnel-coupled Bose-Hubbard chains strong interactions can be achieved by quenching the kinetic energy of the atoms. Thereby, Luttinger parameters $K$ of order one, are reachable when considering non-integer fillings, which avoids the Mott insulating regimes. To this end, we will focus in our numerical simulations presented below on the regime of infinitely strong interactions, corresponding to $K=1$. Deviations from this regime, will lead to changes in the microscopic parameters of the field theory. However, the qualitative behavior will remain.
In the infinitely repulsive regime, the Bose-Hubbard ladder maps to coupled Heisenberg chains. In such spin chains the regime of validity of the sine-Gordon description has been thoroughly analyzed in Ref. \cite{Wybo2022}.

\section{Controlled soliton imprinting}
\label{sec_solitons}

Creating, manipulating, and detecting the fundamental excitations of an emergent sine-Gordon field theory is a challenge. One possibility is acting on the microscopic model with an inhomogeneous perturbation such that the ground state of the emergent field theory hosts initially localized solitons. Proposals for achieving this are based on Raman-coupled quasi-condensates \cite{Kasper2020}, and motivated further theoretical analysis of the propagation of solitons \cite{chelpanova2022,Horvath2022}. By contrast, for the experimental realization we are proposing, the goal is to prepare the ground state of the field theory, and subsequently to dynamically imprint the soliton.

In this section, we propose the protocol for imprinting a soliton on the exact ground state of the two tunnel-coupled Bose-Hubbard chains. Protocols for approximately realizing the ground state are discussed in Section~\ref{sec:exp_re}.
Our strategy aims at carefully tuning the spatial profile of the relative phase $\phi(x)$ such that it undergoes a $2\pi-$phase slip when traversing the system.
In order to avoid a massive creation of excitations during the preparation, smooth and slow parameter changes are paramount. 

We begin by introducing a space-time dependent chemical-potential gradient between the two chains $H\rightarrow H +  \sum_j \Delta_j(\tau) (n_{j\uparrow}-n_{j\downarrow})$ that induces a relative phase drift between the two halves of the system~\cite{Pigneur2018}. 
This is best seen from bosonizing the density gradient $(n_{j\uparrow}-n_{j\downarrow})\to \Pi(x)$. The sine-Gordon Hamiltonian thus gets an additional term
$ H^{\text{SG}} \to H^{\text{SG}} +\int \dd x\, \Delta(\tau,x)\Pi(x)$ leading to a deformed equation of motion $\partial_t\phi=\frac{ v_s \pi}{K}\Pi+\Delta(\tau,x)$, where  $\Delta(\tau,x)$ denotes the potential imbalance in the continuum limit.
The gradient $\Delta(\tau,x)$ hence acts as a source for the accumulation of the relative phase $\phi$. When winding up the phase in half of the system, while leaving the other untouched, a relative phase difference of $2\pi$ can be achieved. This configuration is then stabilized by the cosine potential of the relative phase in the sine-Gordon model \eqref{eq_H_SG}. At this point, the potential imbalance needs to be switched off in the preparation protocol.
To prevent the phase to slip into the next minimum or to strongly oscillate, we simultaneously deepen the potential barrier by increasing the inter chain hopping $t_\perp$, which helps pinning down the phase to the desired value. The result of our protocol is the creation of a localized wavepacket of excitations, primarily containing the desired number of solitons. The wavepacket is approximately contained within the regions where the chemical potential gradient $\partial_x \Delta(\tau,x)$ is not zero.
A graphical summary of the soliton creation protocol can be found in \figc{fig:sol_curr}{b-c}. 

Within bosonization the rung current is proportional to the sine of the relative phase: {$J_j\equiv -i b^\dagger_{j \uparrow}b_{j \downarrow}+i b^\dagger_{j \downarrow}b_{j \uparrow}\simeq 2 |\gamma|^2 \sin\phi$}. We can thus optimize the parameters of the protocol by measuring the current between the two rungs and requiring a smooth sine-shaped profile without additional oscillations [see \figc{fig:sol_curr}{d,e}]. The data is obtained numerically using the time-evolving block decimation (TEBD) algorithm~\cite{Vidal2004} for two coupled Bose-Hubbard chains of length $L=128$ at filling $n=1/8$ and Luttinger parameter $K=1$. In Appendix~\ref{app_different_dentisities}, we show the current profiles for different filling fractions. Depending on the strength of the chemical-potential gradient, we can also control the number of zero-crossings in the current profile. Therefore, our data suggests that a quantized number of solitons can be created in the quantum limit of the sine-Gordon theory in a controlled way using this protocol. 

We furthermore numerically compute the energy of the state before and after the soliton imprinting. For the data shown in \figc{fig:sol_curr}{d-e}, we find that the additional energy is $\Delta_{1S} = 0.50t$ and $\Delta_{2S} = 0.92t$. This should be compared to the cost of creating a soliton measured by the energy gap. For $K=1$, one has that the dispersion relations of the soliton and the first breather are equal, $E_{B_1} = E_S$. Hence, we can extract the energy of the soliton by measuring the energy of the lowest breather in the spectral function~\cite{Wybo2022}. 
From that we estimate $E_S(k=0) = (0.24 \pm 0.04)t$. Therefore, slightly more energy is pumped into the system by the preparation protocol, which is  reasonable as there will also be excitations of the gapless symmetric modes and furthermore slightly-dispersing wave packets of the solitons are created. Nonetheless, it would be desirable, to have access to other more direct observables to characterize the solitons.

One approach could be to track the phase  from individual snapshots of the current and locate phase jumps therein. This strategy has been successfully applied deep in the semiclassical limit, characterized by a large Luttinger parameter $K$ for coupled quasi-condensates~\cite{Schweigler2017}, where solitons have a large spatial extent and can therefore be directly imaged. 
In the setting considered here, however, the local current can only take three values $+, 0, -$ depending on the configuration of the hardcore bosons on a single rung. Even in the case of finite Hubbard interactions, the local current is still quantized to integer values.

Therefore, coarse-graining over spatially extended regions has to be performed. It is crucial that averaging on scales larger than the soliton itself has to be avoided, because then the phase slip cannot be resolved either.
We estimate the size of the soliton using a classical soliton profile that at rest obeys $\ell\partial_x\phi= \sqrt{1-\cos\phi}$, with $\ell =2\hbar K/(\pi v_s M)$. For $K = 1$ and $n=1/8$, we obtain $Mv_s^2\simeq 0.25 t$, giving $\ell\simeq 2$ lattice sites. 
From these estimates, it can be deduced that the phase profile of the soliton cannot be directly extracted, as coarse-graining is required to resolve the phase, but the soliton itself is very localized in space. This is a direct consequence of being deep in the quantum regime of the sine-Gordon model.

To confirm this picture we perform numerically projective measurements of the current: upon coarse-graining all phase slips are averaged out rapidly, see Appendix~\ref{app_current_snap}. 
Due to the small size of the classical soliton one may question the validity of the field theory description. However, it should be emphasized that the classical estimate is only a rough qualitative indicator. Indeed, a careful quantitative study of coupled spin-chains~\cite{Wybo2022} shows that the sine-Gordon model is indeed a faithful description of strongly-interacting ladder systems.
As an alternative to characterize and count the solitons we propose instead to investigate the topological charge.

\section{Detecting solitons from the topological charge}
\label{sec_soliton_detection}

Among the infinite set of conservation laws of the sine-Gordon model, that arise from integrability \cite{korepin1997,Smirnov1992}, is the topological charge, which is an integral over a total derivative,
\be
    Q = \int \frac{\dd x}{2\pi} \,  \partial_x\phi(x)\, .
\ee
Using the canonical commutation relations $[\phi(x),\Pi(y)]=i\delta(x-y)$ one can readily obtain $[Q,H^\text{SG}]=0$. The topological charge is quantized in integer units: each soliton contributes $+1$, anti-soliton $-1$, and breathers, which are bound states of a soliton and an anti-soliton, do not contribute at all. For our analysis it will be useful to introduce the accumulated topological charge $Q(x)=\int^x \frac{\dd y}{2\pi}\partial_y\phi$, which is only conserved (and quantized) as $x\to\infty$. The accumulated topological charge has the advantage that it is automatically coarse-grained over a large portion of the system.

\begin{figure}
    \centering
    \includegraphics[width=0.49 \textwidth]{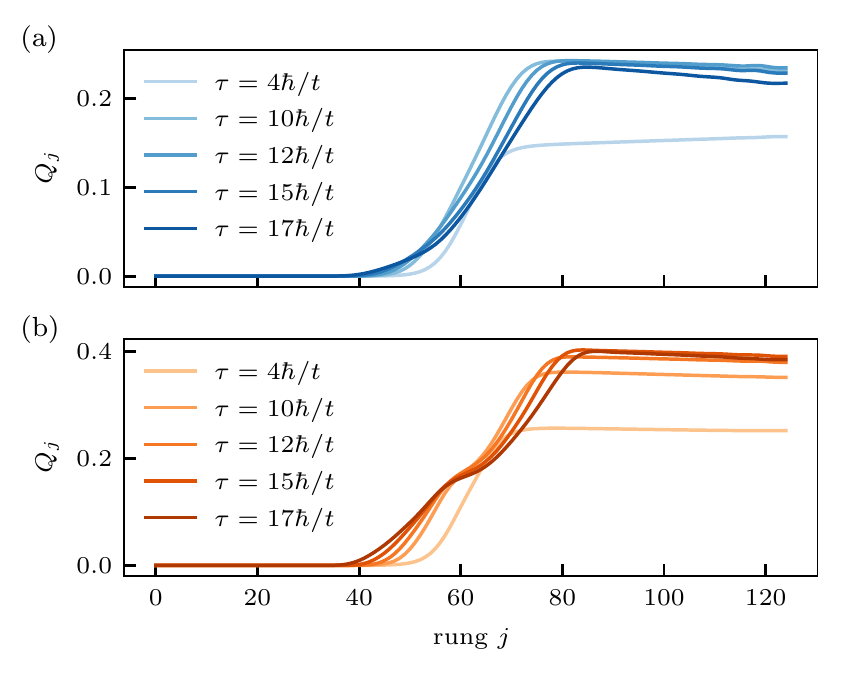}
    \caption{\textbf{Accumulated topological charge.} 
    The accumulated topological charge is shown for one-soliton (a) and two solitons (b) at various times (legend). The soliton creation protocol and system size are the same as in \fig{fig:sol_curr} and takes a total time of $10 \hbar/t$. 
    Thus for the earliest times shown, the imprinting has not yet completed. Once the solitons have been created, the total topological charge remains remarkably constant. 
    The lattice regularization of the topological charge is not quantized to unity, due to the non-universal prefactor $\tilde \gamma$ in Eq. \eqref{accuZ}.
    However, the topological charge for the two-solitons (b) is approximately twice the one of the single soliton (a), indicating a quantization of the excitation.}
    \label{fig:topo_tslice}
\end{figure}

Our goal is now to find a microscopic realization of the accumulated topological charge.
To this end, it is useful to  introduce the local current {$J_j =(-ib^{\dag}_{j\uparrow } b_{j\downarrow } + \mathrm{h.c.})$ }
and energy $E_j = (b^{\dag}_{j\uparrow } b_{j\downarrow } + \mathrm{h.c.})$ operators acting on a single rung.
Using bosonization, the topological charge density can be identified as the most relevant contribution to the following combined observable
\be \label{lattice_charge}
E_j J_{j+1}-J_{j-1} E_{j+2}\simeq 8|\tilde{\gamma}|^4 \partial_x\phi\, .
\ee
The energy operators are measured on even rungs and currents on odd ones and  $\tilde{\gamma}\simeq \gamma$; a possible difference between the two may arise due to UV field theory renormalization.
The actual value of $\tilde{\gamma}$ is however unimportant for us.
With this identification, we can readily introduce a lattice version of the accumulated topological charge as
\begin{multline}\label{accuZ}
Q_j\equiv\sum_{i \text{ even}}^{j} (E_i J_{i+1}-J_{i-1} E_{i+2})\simeq 16\pi|\tilde{\gamma}|^4 Q(x)= \\= 16\pi|\tilde{\gamma}|^4 [\phi(x)-\phi(0)]\, .
\end{multline}
We numerically evaluate this lattice regularization of the topological charge for the state with one and two solitons, respectively, discussed in the previous section; see Fig.~\ref{fig:topo_tslice}. 
The accumulated topological charge captures the phase difference between a given site in our system and the left boundary. The chemical potential gradient $\Delta_j$ has been chosen to act on the right half of the system. Therefore, the left boundary does not evolve and $\phi(0)$ is a mere constant. As a consequence, $Q_j$ probes the dynamics of the field $\phi(x)$.
We find that during the soliton creation process ($\tau < 10 \hbar/t$) the value of the topological charge is still drifting.  However, as the soliton creation process has finished, the topological charge remains remarkably stable in time, as demonstrated by the plateau in Fig. \ref{fig:topo_tslice}.
By comparing the one-soliton and two-soliton imprinting, respectively shown in Fig.~\ref{fig:topo_tslice}(a) and Fig.~\ref{fig:topo_tslice}(b), we observe the total phase slip of the two-soliton state is about twice the one of the one-soliton state, as expected.
The stability of the soliton is a clear signature of the emergent sine-Gordon field theory, because the accumulated charge is just a sum of local energy and current operators without any evident topological property. Thus the presence of the underlying soliton is quite remarkable. 

The topological charge density as defined in Eq.~\eqref{lattice_charge} can be directly measured with Quantum Gas Microscopes \cite{Gross2021}. To this end, a proper local basis rotation has to be performed with beam-splitting operations on individual rungs of the Bose-Hubbard ladder. 
The simpler operator is the current, which can be obtained as follows \cite{Atala2014}. 
After the hopping between the rungs is frozen by strongly increasing the depth of the optical lattice, each of the individual rungs are evolved in time for a duration of $\pi/4$. This realizes a unitary $U_j^J=\exp(i \frac{\pi}{4} E_j)$ which transforms the relative density to the rung current

\be \label{eq:rot1}
(U^{J}_j)^{\dag}  (n_{j \downarrow } - n_{j \uparrow }) U^{J}_j =  J_j \, .
\ee
Here, crucially, we assume the absence of double occupancies. In the case of infinite repulsion, which we consider, this is trivially the case. However, also in the experimentally accessible limit of strong but finite interactions real double occupancies can be neglected, because we are focusing on the dilute limit and strong interactions. As an alternative, arbitrary occupancies of the double well will be allowed if one switches off the interactions during the rotation for example by Feshbach resonances.

In a similar way, the rung energy $E_j$ can be transformed to the relative density, by first applying a potential imbalance, which couples to the density difference, for period $\pi/4$ and then applying the beam splitter operation, leading to  $U_j^E=U_j^J \exp[i\frac{\pi}{4}(n_{j \uparrow}-n_{j \downarrow})]$. 
This transformation maps the density difference to the local energy density
\be \label{eq:rot2}
  (U^{E}_j)^{\dag}  (n_{j \uparrow } - n_{j \downarrow }) U^{E}_j =  E_j \, .
\ee

\begin{figure}[t!]
    \centering
    \includegraphics[width=0.48\textwidth]{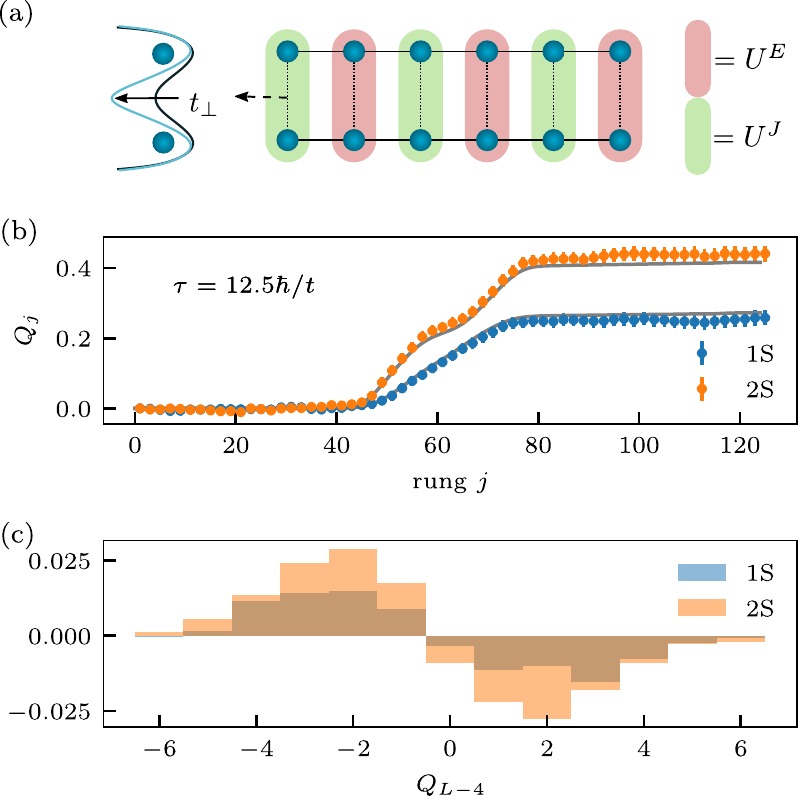}
    \caption{\textbf{Snapshots of the topological charge.} (a) After freezing the hopping within the chain, a density gradient is applied for a period of $\pi/4$ to even sites, pink, followed by a global beam-splitter tunneling operation for a time $\pi/4$. This is effectively described by the unitaries $U^{J}$ and $U^{E}$. (b) The accumulated topological charge extracted from the snapshots for the one- and two-soliton state {prepared as in Fig.~\ref{fig:sol_curr}}. The average is taken over $\approx 8000$ snapshots. The errorbars represent the standard error of the mean, and the grey line is the expectation value of the state. (c) The differential histograms of the total topological charge between the soliton states and the ground state. Deep in the quantum regime, the soliton manifests itself in the skewness of this distribution.  }
    \label{fig:snapshots}
\end{figure}

By first applying a potential gradient on every even rung and then applying the beam-splitter operation $U^J$ globally, as depicted in Fig.~\ref{fig:snapshots}(a), projective measurements in the density basis yield the staggered string operators `$J_0E_1J_2E_3J_4E_5...$'. From those the accumulated charge $Q_j$ \eqref{accuZ} is then directly obtained. By numerically sampling snapshots after performing the beam-splitter operations~\cite{Ferris2012,Buser2022}, we generate the full counting statistics of the topological charge; see Appendix~\ref{app:topoSnapshots} for details. The average over the accumulated topological charge over 8000 snapshots are shown in Fig.~\ref{fig:snapshots}(b), where we find good agreement with the direct measurement of the state.
Moreover, we show a difference of the full distribution function of the topological charge $Q_{L-4}$ for a state in which solitons are prepared compared to the ground state which does not carry any solitonic excitations; Fig.~\ref{fig:snapshots}{(c)}. We note that the $x$-axis of this histogram is quantized to integer values, as $Q_j$ is an integer when it is taken from a single snapshot. Deep in the quantum regime, the soliton manifests itself as a skewness in the broad distribution of $Q_j$. Deep in the classical regime, by contrast, the full distribution function of the topological charge would exhibit a sharp peak.

\section{Ground state preparation} \label{sec:exp_re}

So far we have assumed that the ground state of the coupled chain has already been prepared. Here, we suggest an adiabatic ground state preparation protocol that can be implemented using ultracold atoms in optical lattices with tunable local potentials. The finite mass gap of the sine-Gordon field theory, and the decoupling between the symmetric and anti-symmetric sector at low energies, enables us to prepare the sine-Gordon ground state in reasonable time. Of course, this only remains true as long as the sine-Gordon model is a good description of the coupled chains, which may not be always the case during the state-preparation process. 

\begin{figure}
    \centering
    \includegraphics[width=0.48\textwidth]{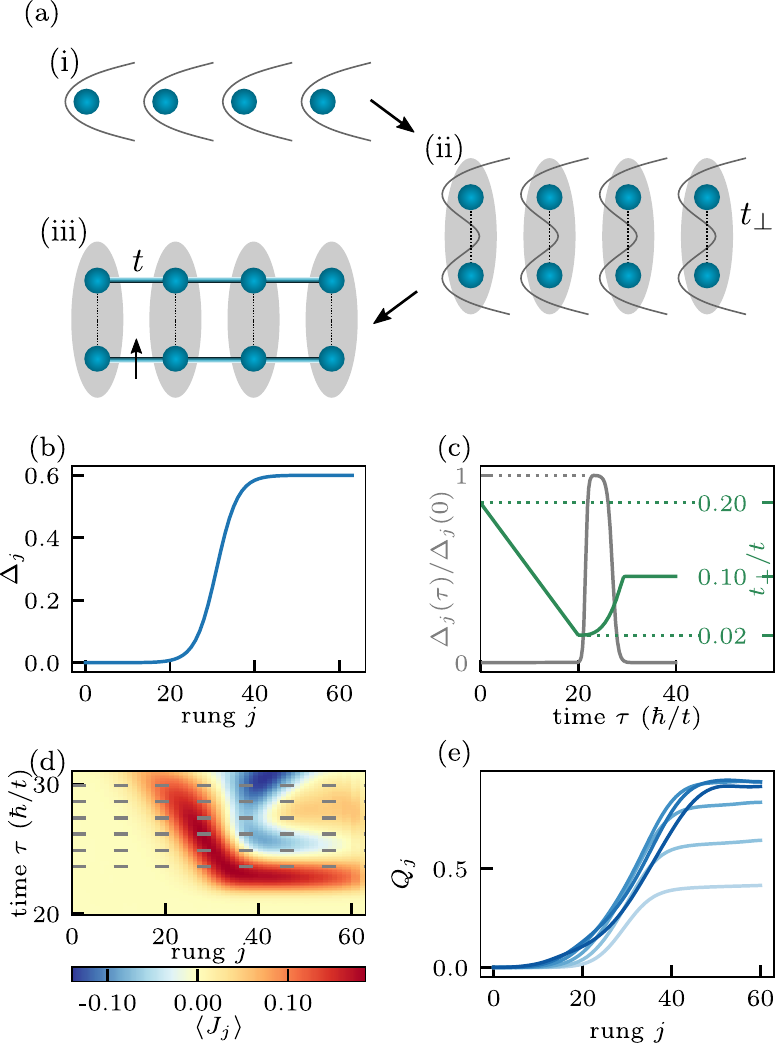}
    \caption{\textbf{Adiabatic ground state preparation.} (a) The ground state preparation protocol starts out by a one-dimensional Mott insulator with negligible hopping, $t \approx 0$. Then the tube is split adiabatically into two,  realizing the ground state on each of the single rungs. Now correlations build up across the system by adiabatically ramping up the intra-chain hopping $t$ over a time $50\hbar/t$, such that after completing the process $t_\perp/t = 0.2$. (b,c) The subsequent soliton imprinting protocol is closely related to the one discussed previously. Before applying the chemical potential gradient profile (b), however, we first lower $t_{\perp}/t$ from $0.2$ to $0.02$ (c). The grey step-like function then shows the time interval during which the relative phase is accumulated. (d) Current profiles around the soliton imprinting stage. (e) Accumulated topological charge shown at time-slices indicated by the grey dashed lines in (d). The lightest shade represents the earliest time and the darkest one the latest.   }
    \label{fig:sol_prep}
\end{figure}

We propose the following adiabatic ground state protocol for creating an approximate ground state of the two tunnel-coupled Bose-Hubbard chains at filling $n=1/2$:
\begin{enumerate}[label=(\roman*)]
    \item Initialize a single chain with unit filling while the intra-chain hopping is switched off. 
    \item Split the potential on every site to prepare the ground state of a single particle in a double well.
    \item Linearly ramp up the intra-chain hopping $t$ to spread correlations through the system.
\end{enumerate}
We simulate this ground state preparation protocol numerically using matrix product states; \fig{fig:sol_prep}(a). To this end, we start out from the ground state of each rung and then increase the intra-chain hopping $t$ over a duration of $50 \hbar/t$, such that the ratio of the inter- and intra-chain hopping is $t_\perp/t = 0.2$ at the end of the protocol. We have not attempted to optimize these numbers too carefully. Rather this should be seen as a guide for the approximate requirements of the adiabatic state preparation protocol. 

The whole system is gapless due to the symmetric degrees of freedom. Hence a large amount of energy is injected in the system even for slow protocols.
However, the symmetric modes are expected to quickly decouple from the sine-Gordon Hamiltonian, which in turn has a finite mass gap. Therefore, the antisymmetric sector remains overall close to its ground state. The total energy it is thus not a good indicator of the quality of the sine-Gordon's state preparation, due to the large contribution from the symmetric modes.

In order to assess the quality of the adiabatiacally prepared state, it can for example be probed spectroscopically, see Appendix \ref{app_BS_spectroscopy}. We numerically find that the low-energy excitations as measured by the spectral function are in good agreement with the spectral function evaluated from the exact ground state. This demonstrates that the asymmetric sector is well prepared using this procedure.

In addition, we can even imprint a soliton on top of the adiabatically prepared ground state, following our suggestion discussed earlier in Section \ref{sec_solitons}; 
for details see \figc{fig:sol_prep}{b,c}. 
The resulting current profiles are again indicative of having a soliton imprinted (d). This is furthermore supported by a reasonably stable accumulated topological charge (e). This demonstrates that the adiabatically prepared ground state has the essential features of the ground state of the emergent sine-Gordon field theory.

\section{Conclusion \& Outlook}
\label{sec_conclusions}

The sine-Gordon field theory emerges as the effective description in a wealth of physical systems. Here, we propose access to the deep quantum regime of the sine-Gordon model by realizing tunnel coupled Bose-Hubbard chains. We discuss adiabatic ground state preparation protocols, as well as protocols for creating the fundamental excitations -- the solitons -- of the sine-Gordon model. The soliton can be most directly seen in the inter-chain current and the topological charge, which both can be measured with quantum gas microscopes using beam-splitter operations.

For future work, it would be interesting to explore scattering of solitonic wave packets. Fundamental information of the excitations and their mutual interactions can be extracted from scattering events. Information about the integrable field theory can thereby be obtained, in which multi-particle scattering processes are rather peculiar \cite{Zamolodchikov1979, Smirnov1992, Wybo2022}.
Moreover, it would be exciting to find ways for creating initial states with a finite density of solitons. In such a true many-body regime  thermalization dynamics~\cite{Calabrese_2016} and the emergent generalized hydrodynamics~\cite{Bertini2016,alvaredo2016,specialissueGHD} of the strongly interacting sine-Gordon field theory could be explored.

\section*{Acknowledgements}

We thank F. Essler for useful discussions. Matrix product state simulations were performed using the TeNPy package~\cite{Hauschild2018}. We acknowledge support from the Deutsche Forschungsgemeinschaft (DFG, German Research Foundation) under Germany’s Excellence Strategy--EXC--2111--390814868 and DFG grants No. KN1254/1-2, KN1254/2-1, the European Research Council (ERC) under the European Union’s Horizon 2020 research and innovation programme (Grant Agreement No. 851161), as well as the Munich Quantum Valley, which is supported by the Bavarian state government with funds from the Hightech Agenda Bayern Plus.

{\par\textit{Data availability:}} Raw data and data analysis are available on Zenodo~\cite{wybo_elisabeth_2023_7778949}.

\clearpage

\onecolumngrid 
\appendix

\section{Soliton profiles at different densities}\label{app_different_dentisities}
\begin{figure*}[t!]
    \centering
    \includegraphics[width=0.95\textwidth]{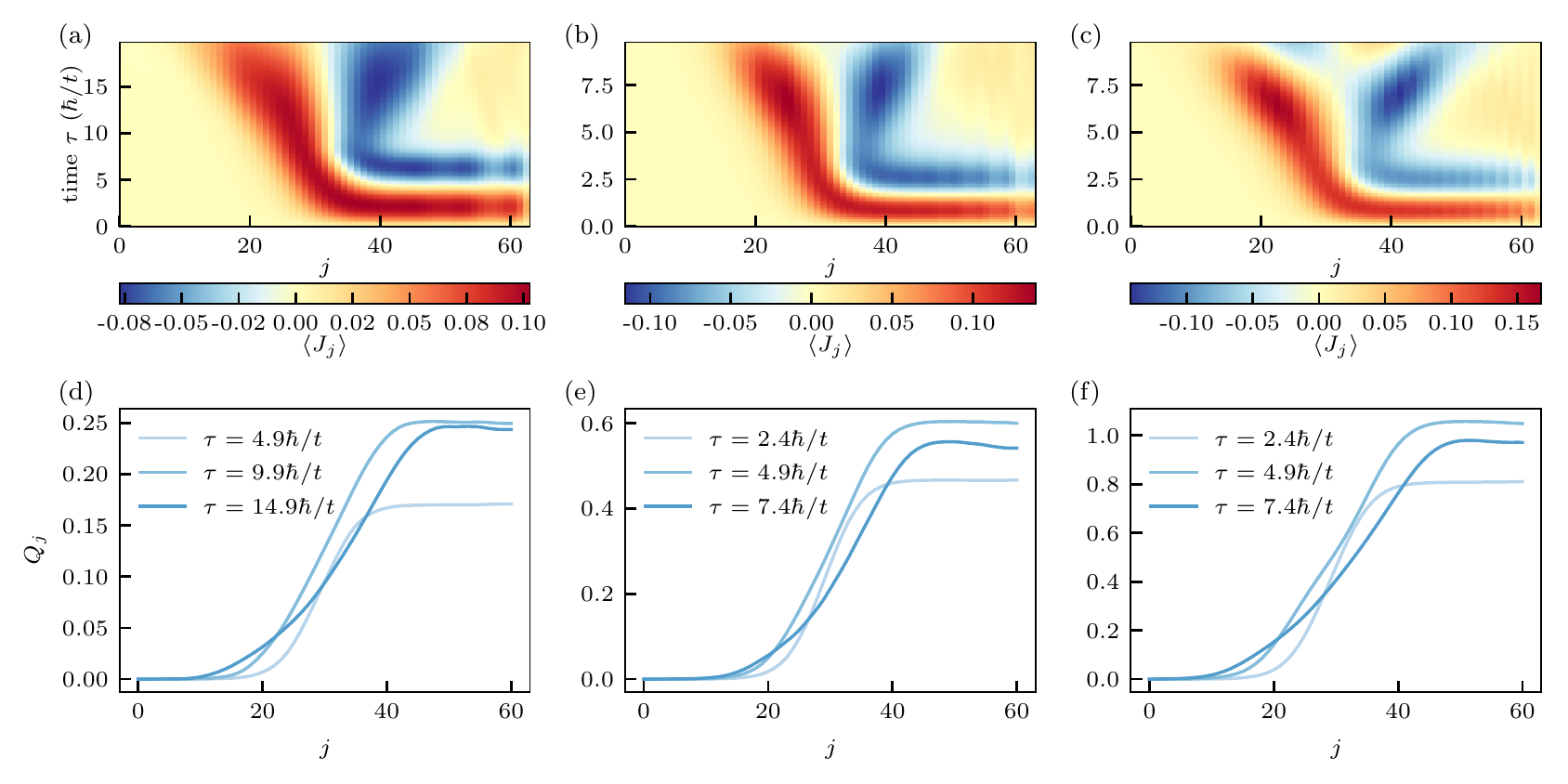}
    \caption{\textbf{Solitons at different densities.} (a-c) The current profiles when applying the soliton creation protocol for different densities in the initial ground state of $L=64$ rungs. (a,d) $n=1/8$, (b,e) $n=1/4$ and (c,f) $n=1/2$. The soliton spreading is significantly faster for higher densities due to the larger sound velocity. (d-f) The corresponding accumulated topological charge for some time slices.   }
    \label{fig:Z_dens}
\end{figure*}

We compare the soliton profiles and the accumulated topological charge obtained for states at different densities; Fig.~\ref{fig:Z_dens}. Apart from tuning the strength of the chemical potential, we also reduce the imprinting time with increasing density. Concretely, the time scales for imprinting are chosen inversely proportional to the density to compensate for the higher sound velocity in systems with higher density. We also numerically find that for a higher density the soliton is more dispersive; Fig.~\ref{fig:Z_dens}. 

We also evaluate the accumulated topological charge, which depends approximately linearly on the density; Fig.~\ref{fig:Z_dens}(d--f). 
There are small drifts of the topological charge in time, signaling a weak breaking of the conservation law of the total topological charge. As previously pointed out, the conservation of the topological charge is not directly implemented in the coupled Hubbard chain, but emerges from the effective field theory. This furthermore confirms that the parameter regime of Fig. \ref{fig:topo_tslice} realizes a high-quality sine-Gordon soliton, while more important corrections beyond sine-Gordon are present at higher densities, as shown in Fig.~\ref{fig:Z_dens}. One reason for the worse stability is the reduced preparation time which is required due to the increased sound velocity, while keeping the system size fixed. It would be interesting to optimize the soliton imprinting to further reduce these undesired effects in future work.

\section{Attempting to extract the phase of the soliton from current snapshots}
\label{app_current_snap}

\begin{figure}[t!]
    \centering
    \includegraphics[width=0.49\textwidth]{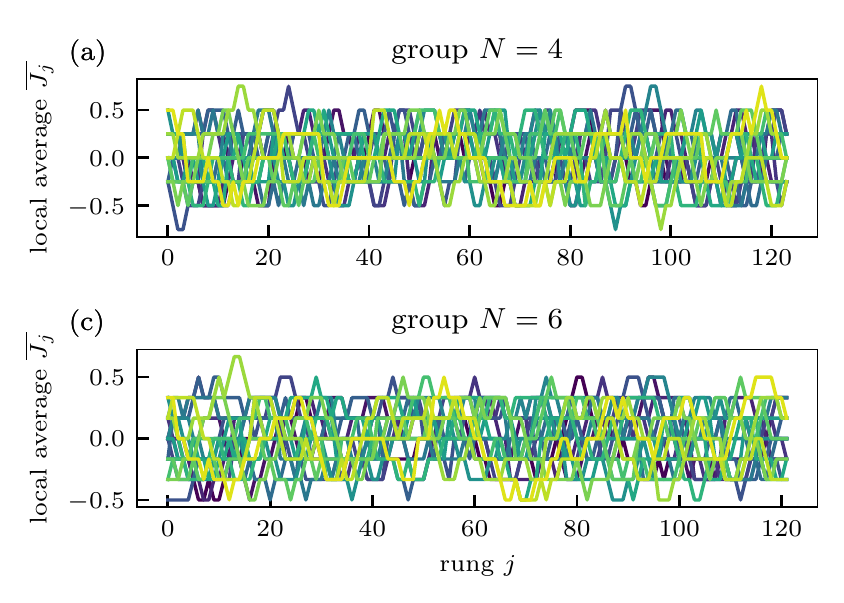}
    \includegraphics[width=0.49\textwidth]{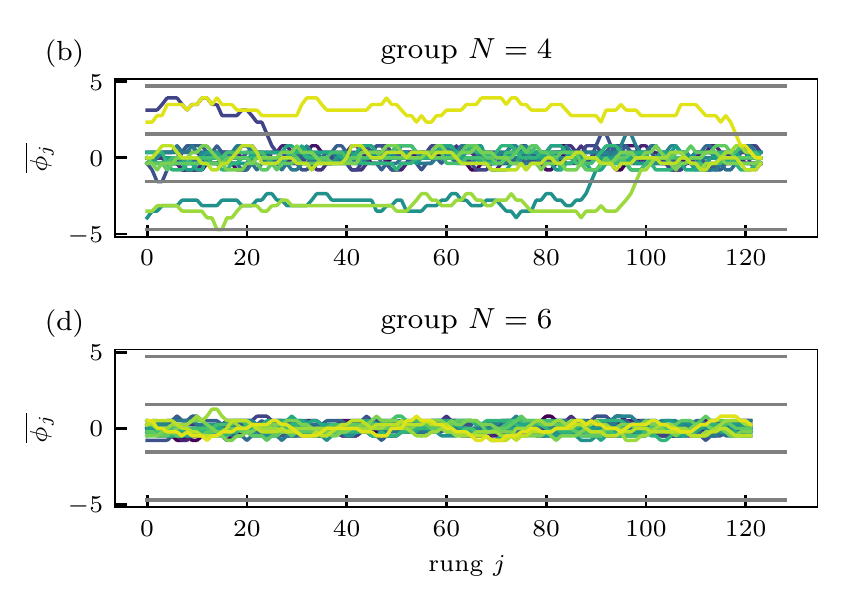}
    \caption{\textbf{Snapshots of the coarse-grained current.} (a,c) Illustration of 20 snapshots of the coarse-grained current for respectively $N=4$ and $N=6$ grouped sites of the 1 soliton state. There are huge fluctuations in this data because we are considering the deep quantum regime of the emergent sine-Gordon theory. (b,d) The relative phase extracted from the snapshots of the current. The rare jumps that are present for $N=4$ disappear when taking $N=6$. The horizontal grey lines are drawn at $\pm \pi/2$ and $\pm 3\pi/2$. The snapshots have been obtained from the one soliton state, presented in the main text in Fig.~\ref{fig:sol_curr} at time $\tau = 12.5 \hbar/t$. }
    \label{fig:JJ_snapshots}
\end{figure}

In this section directly analyze the snapshots of the current {$J_j =(-ib^{\dag}_{j\uparrow}b_{j\downarrow} + \mathrm{h.c.}) $} and show that deep in the quantum regime one cannot extract the phase directly upon coarse-graining. As we discuss in the main text, the reason is that in this regime the soliton itself is of very small extent and hence is averaged out upon coarse-graining. 
Let us describe the formal procedure here. Within bosonization, the current operator can be written in terms of the phase field as
\begin{equation} \label{eq:curr_phase}
J \simeq 2|\gamma|^2 \sin(\phi),
\end{equation}
where $\gamma$ is a non-universal prefactor. When taking snapshots of the current $JJJJ\dots$, it is conceivable in principle that one can extract the phase profile in space. The challenges are the following: first a reliable estimate of the non-universal prefactor $\gamma$ is needed. Secondly (and more importantly) coarse graining has to be performed to obtain the phase. A snapshot of the current on the ladder takes the values
{\begin{equation}
    N_{j\downarrow} - N_{j\uparrow} \in\{ -1,0,1\}\, ,
\end{equation}}
where $N_{j\alpha} = 0,1$ is the measurement outcome of $n_{j\alpha}$. We do not consider higher occupancy as we are in the limit $K\rightarrow 1$ and at low average filling. Then we coarse-grain the current over a certain number of sites, and invert the relation~\eqref{eq:curr_phase} while flattening the grouped current with the value of $2|\gamma|^2$. We extracted a value of $2|\gamma|^2\approx 0.7$ from the spectral function, however, the exact value does not considerably affect the outcome of the procedure. There is always an ambiguity in the inversion of~\eqref{eq:curr_phase}, we choose to make a `jump' in the phase when the absolute value of the current hits it maximum (here taken to be 0.7) once. Another maximum with the same sign corresponds to a jump back. In Fig.~\ref{fig:JJ_snapshots}{(a,c)}, we show 20 snapshots of the coarse-grained current
{\begin{equation}
    \overline{J_j} = \frac{1}{N} \sum_{i=0}^{N-1}  (N_{i+j\downarrow} - N_{i+j\uparrow}) 
\end{equation}}
over respectively $N=4$ and $N=6$ sites. In Fig.~\ref{fig:JJ_snapshots}{(b,d)}, we then show show the extracted phase 
\begin{equation}
    \overline{\phi_j} = \arcsin \left( \frac{ \overline{J_j}}{2|\gamma|^2} \right),
\end{equation}
where $\overline{J_j}$ is thus flattened to $2|\gamma|^2$ in case $\overline{J_j}>2|\gamma|^2$. From this it is already clear that the rare phase jumps, completely disappear upon coarse graining already over a couple of sites. In the random subset of 20 snapshots that are shown, there is not even a single one that winds up a phase of $2\pi$ according to our procedure when grouping $N=6$ sites.

\begin{figure}
    \centering
    \includegraphics[width=0.49\textwidth]{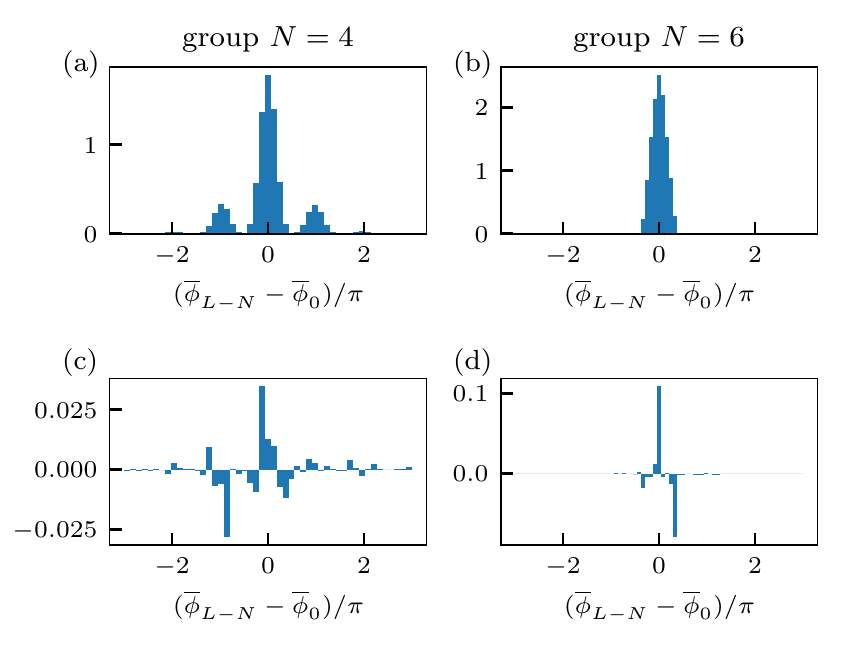}
    \caption{\textbf{Histograms.} (a,b) Histograms of the phase accumulation across the system obtained from $\approx 40\ 000$ snapshots of the one soliton state for respectively $N=4$ and $N=6$ grouped sites. (c,d) The differential histograms comparing the soliton data to the ground state.    }
    \label{fig:JJ_histograms}
\end{figure}

In order to have a larger sample size of snapshots, we create a histogram of the phase accumulation obtained in around $40 000$ snapshots; Fig.~\ref{fig:JJ_histograms}{(a,b)}. These histograms are symmetric, and do not show special features. To strengthen this point, we plot the differential histograms between the ground state and the state containing one soliton; Fig.~\ref{fig:JJ_histograms}{(c,d)}. Except for statistical fluctuations we find that there is no difference between these states. This illustrates that that no special features can be obtained from the coarse grained data. Taking the estimated size of the soliton of two lattice sites, which we give in the main text, this finding is not surprising. It just reflects that the regime we are considering is dominated by quantum fluctuations. 

\section{Extracting the topological charge from snapshots \label{app:topoSnapshots}}

\begin{figure*}
    \centering
    \includegraphics[width=0.95\textwidth]{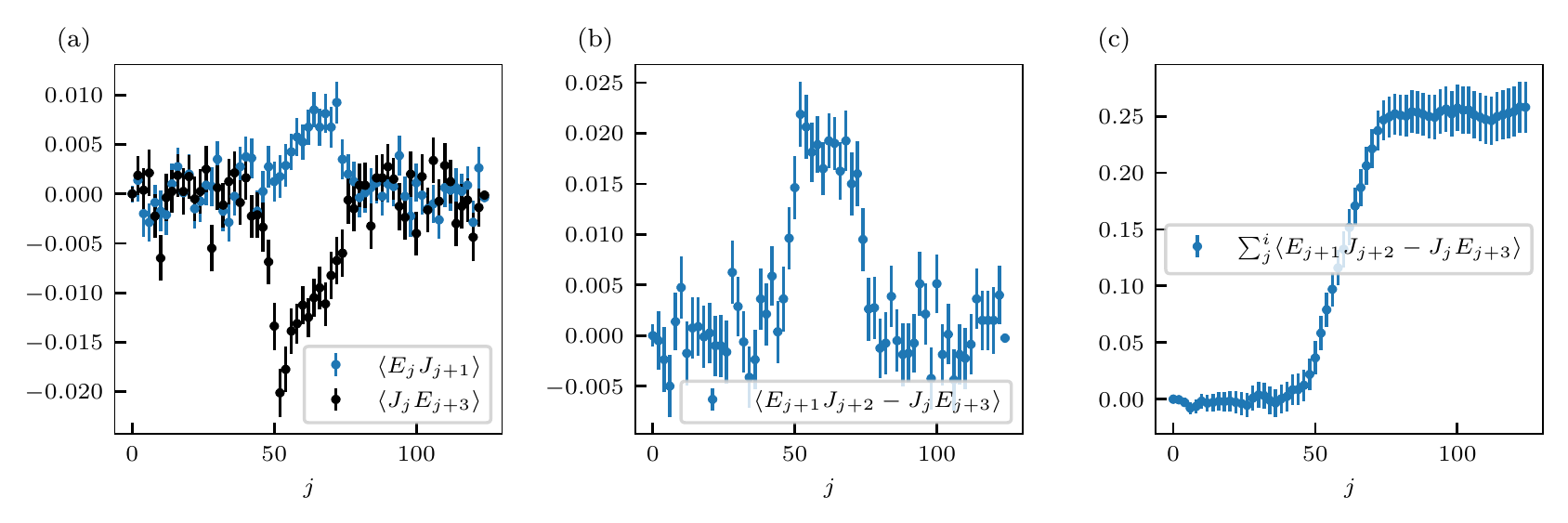}
    \caption{\textbf{Snapshots for determining the topological charge.} (a) The average data extracted from the snapshots. For an odd rung index $j$, we show $E_iJ_{i+1}$, and for even rung index $j$, we show $J_jE_{i+3}$. These expectation values deviate from zero around the position of the soliton. The averages are taken over $N\approx 8000$ snapshots. (b) From the difference between the two expectation values, we compute the topological charge density. (c) The cumulative profile of it then gives the  topological charge $Q_j$.  }
    \label{fig:Z_const}
\end{figure*}

In this section, we show more in detail how to extract the soliton profile from snapshots of the topological charge density. After performing the rotations corresponding to $U^J$ and $U^E$, see respectively Eqs.~\eqref{eq:rot1} and~\eqref{eq:rot2} in the main text, and taking the snapshots in that basis $JEJE\dots$, we can  reconstruct the expectation value of $\ev{J_jE_{j+3}}$ (for even $j$) and $\ev{E_{j}J_{j+1}}$ (for odd $j$). This is shown in Fig.~\ref{fig:Z_const}{(a)}. These expectation values deviate from zero around the position of the soliton. Taking then the difference corresponds to the topological charge density which exhibits a bump at the position of the soliton; Fig.~\ref{fig:Z_const}{(b)}. Taking the cumulative sum then reveals the accumulated topological charge, which is quantized when traversing the full system; Fig.~\ref{fig:Z_const}{(c)}.  

\section{Breather spectroscopy of the adiabatically prepared state}
\label{app_BS_spectroscopy}

\begin{figure}
    \centering
    \includegraphics[width=0.49\textwidth]{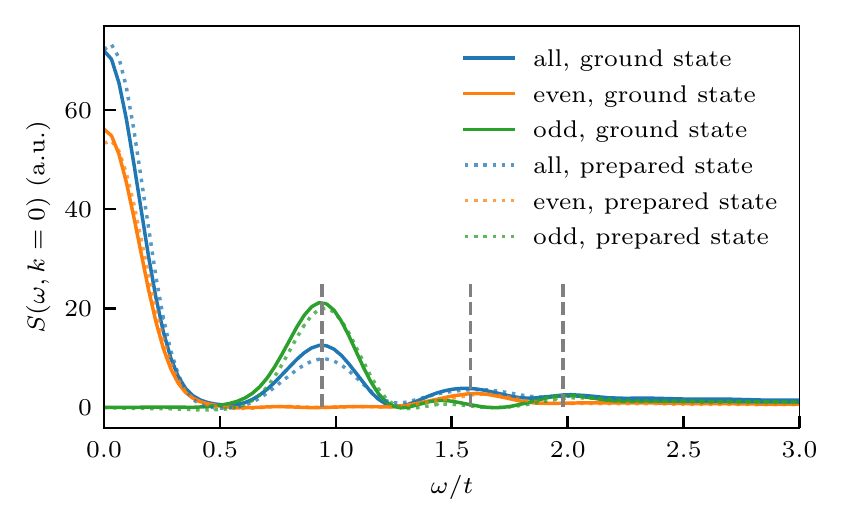}
    \caption{\textbf{Breather spectroscopy.} The spectral function at zero momentum, $k=0$, as defined in Eq.~\eqref{eq:spec_f} is computed for the ground state and the adiabatically prepared ground state at $t_{\perp}/t=0.2$. This state is prepared as described in the main text for a system of $L=32$ rungs within a time of $50\hbar/t$. The spectrum is computed separately for the different parity sectors of the breather. The grey dashed lines show the location of the exact energies of the three breathers present for $K=1$. The spectra are in very good agreement, demonstrating that the ground state of the anti-symmetric sector is well prepared by our procedure.  }
    \label{fig:breather_spec}
\end{figure}

Here, we analyze the breather spectrum of the  adiabatically prepared ground state following the protocol of Sec.~\ref{sec:exp_re}, and compare it to the spectrum of the exact ground state on the coupled chains. To this end, we compute the spectral function at momentum $k=0$
\begin{equation} \label{eq:spec_f}
    S(\omega,k=0) \sim \int \dd t e^{i\omega t} \sum_{n=0}^{L-1} \ev{O_n(t)O_{L/2}}{\tilde{\psi}_0},
\end{equation}
where $\ket{\tilde{\psi}_0}$ is the (adiabatically prepared) ground state. The operators $O_n$ are defined on the rungs, to couple to the asymmetric sector in which the sine-Gordon theory is emerging. We observe a nice correspondence between the spectral functions and  the expected breather energies; Fig.~\ref{fig:breather_spec}. Data is shown for three different operators: (i) $O_i = b^{\dag}_{i\uparrow}b_{i\downarrow}$ coupling to all breathers, (ii) $O_i = b^{\dag}_{i\uparrow}b_{i\downarrow} + \mathrm{h.c.}$ coupling to the even breathers only, and (iii) $O_i = ib^{\dag}_{i\uparrow}b_{i\downarrow} + \mathrm{h.c.}$ coupling to the odd breathers. 
 
\twocolumngrid  
\bibliography{biblio}

\end{document}